\documentclass[aps,prl,groupedaddress,twocolumn]{revtex4}

\usepackage{graphicx}
\usepackage{color}
\usepackage{nicefrac}
\usepackage{amsmath}
\usepackage{amssymb}
\usepackage[pdftex,colorlinks=true,linkcolor=blue,citecolor=blue,filecolor=blue,baseurl=empty]{hyperref}

\newcommand{\avg}[1]{\big< #1 \big>} % for average
\renewcommand{\v}[1]{\ensuremath{\mathbf{#1}}} % for vectors
\newcommand{\ket}[1]{\big| #1 \big>} % for Dirac bras
\newcommand{\matrixel}[3]{\big< #1 \vphantom{#2#3} \big| #2 \big| #3 \vphantom{#1#2} \big>} % for Dirac matrix elementss
\newcommand{\lr}[1]{\left(#1\right)}
\newcommand{\tot}[1]{#1_{\text{tot}}}
\newcommand{\eff}[1]{#1_{\text{eff}}}
\newcommand{\dir}[1]{#1_{\text{dir}}}
\newcommand{\commut}[2]{\left[#1,#2\right]}
\newcommand{\norm}[1]{\left|#1\right|^2}

\newcommand{\ff}[1]{{\boldsymbol #1}}

\begin{document}

\title{Magnetic Doublon Bound States in the Kondo Lattice Model}

\author{Roman Rausch$^{1}$}
\email[]{rausch.roman.72e@st.kyoto-u.ac.jp}
\author{Michael Potthoff$^{2}$}
\author{Norio Kawakami$^{1}$}

\affiliation{$^{1}$Department of Physics, Kyoto University, Kyoto 606-8502, Japan}
\affiliation{$^{2}$Department of Physics, University of Hamburg, Jungiusstra{\ss}e 9, D-20355 Hamburg, Germany}

\begin{abstract}
We present a novel pairing mechanism for electrons, mediated by magnons.
 These paired bound states are termed ``magnetic doublons''. Applying numerically exact techniques (full diagonalization and the density-matrix renormalization group, DMRG) to the Kondo lattice model at strong exchange coupling $J$ for different fillings and magnetic configurations, we demonstrate that magnetic doublon excitations exist as composite objects with very weak dispersion. They are highly stable, support a novel ``inverse'' colossal magnetoresistance and potentially other effects.
\end{abstract}

%\keywords{}

\maketitle

\paragraph{\color{blue} Motivation.}  
The Mott-insulating state of the single-band Hubbard model, driven by repulsive Coulomb interaction, remains an enduring paradigm in strongly-correlated electron physics. 
An excitation of electrons across the Mott gap to the upper Hubbard band forms ``doublons'', quasiparticles which are stabilized by the strong Hubbard interaction $U$ and which persist as bound states on a time scale growing exponentially with $U$ \cite{Strohmaier_2010, Sensarma_2010}. 
This stability provides a bottleneck in the relaxation dynamics \cite{Avigo_2019}, and several propositions have been made to exploit this effect: 
(i) Since doublons effectively behave as hard-core bosons with an attractive interaction, the creation of a metastable superfluid state is conceivable. 
Being an excited state far from equilibrium, however, such a state has to be engineered in an optical lattice \cite{Rosch_2008} or by photodoping \cite{Kaneko_2019}. 
(ii) The stability of doublons on a long time scale makes Mott insulators candidate materials for solar cells \cite{AlHassanieh_2008, Eckstein_2014}. 
In this case, incident light creates a doublon-hole excitation which quickly recombines if there is no interaction, leading to merely diffusive charge separation. 
With interaction, on the other hand, a stable doublon is created that can be separated from the hole more efficiently. 
(iii) The so-called ``quantum distillation'' is a cooling method for atoms in optical lattices \cite{Heidrich-Meisner_2009, Xia_2015, Scherg_2018}. 
Having trapped a number of doublons and then releasing the trap, one observes that while fast components of the block escape, the rest bunches together, leading to an increase of the local double occupancy and hence to an approximate band insulator with low entropy.

Here, we demonstrate the existence of a novel quasiparticle, the ``magnetic doublon''. 
Opposed to the conventional one, it is stabilized by magnetic degrees of freedom rather than by a strong  Hubbard-$U$. 
The magnetic doublon is in fact a bound state of a conventional doublon and a magnon. 
It requires a magnetic background and consists of a double occupancy in the vicinity of a spin-flip excitation, thus being a quasiparticle that is itself formed out of two quasiparticles.

We will argue that the Kondo lattice model (KLM) is the most simple system that hosts the magnetic doublon. 
Key to an understanding of the binding mechanism is the limit of a strong local exchange coupling $J$.
This regime is of high relevance for materials like doped manganites \cite{Kubo_1972, Ishihara_1997} and is precisely the limit accessible in the simulation of the KLM by an ultracold $^{173}$Yb gas in an optical lattice \cite{Riegger_2018}. 
As the concept is fairly general, one should expect magnetic doublons in other contexts as well, e.g. in multi-orbital, strong-$J$ systems, such as Hund's metals \cite{YHK11}.
The magnetic doublon provides an alternative route to the same functionalities (i)--(iii) of the conventional doublon but in addition to that also has a couple of further intriguing properties as will be shown here.

\paragraph{\color{blue} Two-electron exact solution.}  
The quintessence of the magnetic doublon concept can be already understood in the two-electron case ($N=2$).
We consider a one-dimensional lattice with periodic boundaries; this is convenient methodically, but not essential. 
The KLM Hamiltonian reads:
\begin{equation}
  H = -T \sum_{\left<ij\right>\sigma} \lr{c^\dagger_{i\sigma} c_{j\sigma} + H.c.} + J\sum_i \v{S}_i \cdot \v{s}_i \: .
\end{equation}
Here, $c^\dagger_{i\sigma}$ creates an electron with the spin projection $\sigma = \uparrow, \downarrow$ at site $i$.
Furthermore, $\v{s}_i = \sum_{\sigma\sigma'} c^{\dagger}_{i\sigma} \ff \tau_{\sigma\sigma'} c_{i\sigma'}/2$ (with the Pauli matrices $\ff \tau$) is the spin of the electron at site $i$ and $\v{S}_i$ is the local spin with quantum number $\nicefrac12$.
$J>0$ denotes the antiferromagnetic (AFM) exchange interaction.
The hopping amplitude $T\equiv 1$ between nearest neighbors (denoted by the angle brackets $\left<ij\right>$) fixes the energy and time units ($\hbar \equiv1$).
The ground state of the model in the strong-$J$ regime on $L$ sites is ferromagnetic (FM) with total spin $S=(L-N)/2$, for $N=1, 2, ..., L-1$ and becomes a singlet $S=0$ at half filling $N=L$ \cite{McCulloch_2001}.

For $N=2$ electrons, i.e. at most one doublon, we can develop a ``doublon band theory'' \cite{Rausch_2016}. Effective doublon-doublon interactions show up at higher fillings; this is discussed later.
We perform full diagonalization to simultaneously get the many-body eigenstates of $H$, the particle-number operator $\hat{N}$, the squared total spin operator $\tot{\v{S}}^2=\lr{\sum_i\lr{\v{S}_i+\v{s}_i}}^2$, and the total momentum operator $\hat{P}$ (derived in \cite{EFG+05}), 
with the corresponding quantum numbers $E$, $N=2$, $S\lr{S+1}$ and $K$, respectively.
Several spin symmetry sectors are possible: 
$S=(L+2)/2$ yields the trivial fully polarized case where the spin-exchange $\propto J$ is inactive.
$S=L/2$, i.e. one spin flip away from full polarization is discussed in the Supplemental Material \cite{suppl}. 
Here, we focus on the case $S=(L-2)/2$ (two spin flips), as this sector also contains the ground state. 

\begin{figure}
\includegraphics[width=\columnwidth]{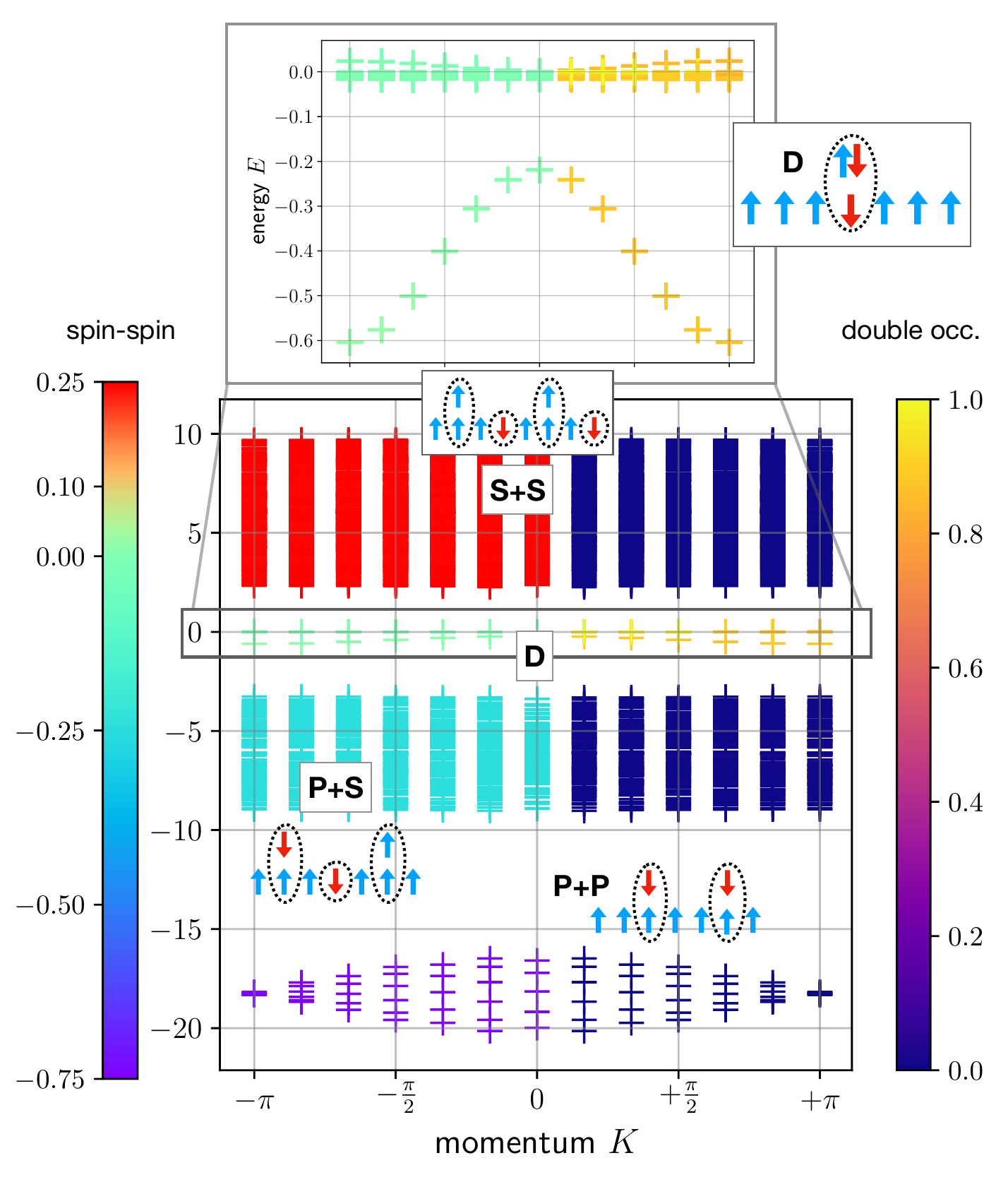}
\caption{\label{fig:eigenstates}
Many-body eigenenergies as a function of the total momentum $K$ of the one-dimensional KLM with $L=12$ sites at $J=12$. 
Each cross refers to an eigenstate in the sector $N=2$, $S=(L-N)/2$. 
The plot is symmetric to $K=0$.
Color code (for $K\leq 0$, left) gives the local spin-spin correlation per electron $\sum_i \avg{\v{S}_i\cdot\v{s}_i}/N$, and (for $K>0$, right) the double occupancy $\sum_{i} \avg{n_{i\uparrow}n_{i\downarrow}}$.
From lowest to highest energy, the states are: \textit{P+P}: two polarons, \textit{P+S}: polaron and scattered electron, \textit{D}: magnetic doublon, \textit{S+S}: two scattered electrons. The doublon bands are magnified in the upper part.
}
\end{figure}

Figure \ref{fig:eigenstates} displays the results. 
Four types of states can be distinguished by calculating the average local spin correlation per electron 
$\langle \v S \cdot \v s \rangle_{\rm loc.} / N \equiv \sum_{i} \avg{\v{S}_i\cdot\v{s}_i} / N$ and the total double occupancy $\avg{n^d} \equiv \sum_{i} \avg{n_{i\uparrow}n_{i\downarrow}}$.
In the lowest-energy states (labeled \textit{P+P}), the electrons are almost completely AFM correlated with the localized spins, so that $\langle \v S \cdot \v s \rangle_{\rm loc.}/2 \approx -\nicefrac34$. 
This causes a fully polarized local-spin system to be favorable because the ``kinetic'' term $\propto T$ of $H$ is minimized.
%by polarizing them towards a ferromagnet. 
For one electron, this limit of a ``magnetic polaron'' has been amply investigated in the past \cite{Nolting_1996, Tsunetsugu_1997, Henning_2012}.
If one of the polarons decays into an FM-aligned electron and a magnon, we get $\langle \v S \cdot \v s \rangle_{\rm loc.} / 2 \approx (-\nicefrac34+\nicefrac14)/2=-\nicefrac14$ (labeled \textit{P+S}).
If both polarons decay, we get the ferromagnetic correlation $\langle \v S \cdot \v s \rangle_{\rm loc.} / 2 \approx + \nicefrac14$ (labeled \textit{S+S}).
However, in between these two scattering ``continua'', around $E=0$, one finds solutions with $\langle \v S \cdot \v s \rangle_{\rm loc.} / 2 \approx 0$ and $\avg{n^d} \approx 1$, which can thus be identified as doublon states (labeled \textit{D}). 
We note that switching the sign of $J$ from AFM to FM simply exchanges the scattering with the polaron part, but does not affect the presence of the doublon in between.

\begin{figure}
\includegraphics[width=\columnwidth]{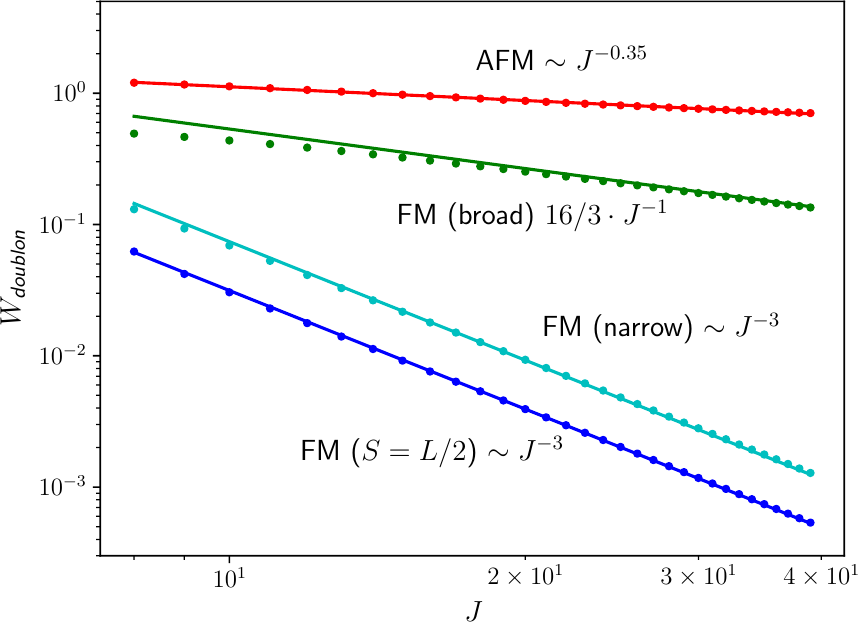}
\caption{\label{fig:W}
Points: Doublon bandwidth as obtained by full diagonalization for various cases. (i)
$S=(L-2)/2$, FM ground state ($L=12$); for broad and narrow dispersions see text.
(ii) $S=L/2$, FM ground state ($L=16$), see Ref.\ \cite{suppl}. 
(iii) $S=0$, AFM ground state ($L=8$), see text.
Lines: Fits $W=\mbox{const} \times J^{-r}$ as indicated, except for the broad doublon bandwidth which is taken from the effective model Eq.\ (\ref{eq:Heff}).
}
\end{figure}

\paragraph{\color{blue} Magnetic doublon.}  
Now, zooming into the doublon states (topmost panel of Fig.\ \ref{fig:eigenstates}), we discover that there are in fact two distinct features with a small gap in between, a narrow and a broad one. 
The narrow one is expected to form a continuum of states in the $L\to \infty$ limit. 
Varying $J$, we find that its bandwidth scales as $W\sim J^{-3}$ for strong $J$ (see Fig. \ref{fig:W}). 
The magnetic doublon corresponds to the broad structure composed of $L$ eigenstates. 
Its dispersion has minima at $K=\pm\pi$ and the bandwidth scales as $W\sim J^{-1}$ (Fig. \ref{fig:W}).

It is possible to analytically derive a first-order effective model that projects onto the states of the broad doublon band by using $T/J$ as a small parameter and neglecting single occupancy (which is essentially the Schrieffer-Wolff transformation, cf.\ Refs.\ \cite{Schrieffer_1966, Tsunetsugu_1997, Rosch_2008} and \cite{suppl}).
We obtain the following result:
\begin{eqnarray}
\eff{H} &=& J_K' \sum_{\left<ij\right>}  \lr{ d^{\dagger}_id_j + d^{\dagger}_jd_i - 2n^d_in^d_j +n^d_i +n^d_j }
\nonumber \\
&\cdot &
\lr{\v{S}_i\cdot\v{S}_j-1/4} \: .
\label{eq:Heff}
\end{eqnarray}
Here $d_i^{\dagger}=c^{\dagger}_{i\downarrow}c^{\dagger}_{i\uparrow}$ creates a double occupancy at site $i$, now a hard-core boson (with $[d_i, d^\dagger_j]=\delta_{ij}$ and $d^{\dagger 2}_i=0$). 
The corresponding density is $n^d_i=d_i^{\dagger}d_i$. 
As compared to the related expression for the Hubbard model \cite{Rosch_2008}, we find a different 
hopping amplitude $J_K' = 8T^2 / 3J$ (rather than $2T^2/U$) and, more importantly, an interaction with the magnetic substrate, given by the spin-spin product in Eq.\ (\ref{eq:Heff}).
Thereby, the sign of the doublon-doublon interaction becomes dependent on the nature of the magnetic ground state. 
It is attractive in the ferro- and repulsive in the antiferromagnetic case.

It is instructive to employ the transformation $d^{\dagger}_i = (-1)^i T^+_i$, $n^d_i = T^z_i + \nicefrac12$  from hard-core bosons to pseudospin operators $\ff T_{i}$, which fulfill the usual spin commutation relations.
We obtain a rather compact spin-pseudospin model:
\begin{equation}
H_{\text{eff}} = -2J_K' \sum_{\left<ij\right>} \lr{\v{T}_i\cdot\v{T}_j-\nicefrac14} \cdot \lr{\v{S}_i\cdot\v{S}_j-\nicefrac14} \: ,
\end{equation}
which is manifestly invariant under the spin- and charge-SU(2) symmetries of the system \cite{Yang_1990, Nishino_1993}.

Note the following important consequence: In the effective model, the subspace $S=(L-2)/2$ contains one doublon and one magnon (the number of states is $L^2$), and $\eff{H}$ only gives a non-zero result if they are at least nearest neighbors. The delocalization of $L$ of these bound states thus forms the ``broad'' dispersion of the magnetic doublon band, while the remaining ones have zero energy.

Numerical evidence shows that the degeneracy of the remaining states is lifted in third order, see Fig.\ \ref{fig:W}, resulting in a narrow continuum where a doublon and a magnon are propagating independently through the lattice.
A scaling of the bandwidth $\sim J^{-3}$ is also seen in the results for the $S=L/2$ sector, see the Supplemental Material \cite{suppl} and Fig.\ \ref{fig:W}.
For $S=L/2$, the phase space is highly restricted and one can either have a doublon or a magnon, but not both. 
Still, the doublon can delocalize in a higher-order process by creating and absorbing a virtual magnon. 
We suspect that the same mechanism is at work for the narrow doublon band of Fig. \ref{fig:eigenstates}.

\paragraph{\color{blue} Inverse CMR effect.}
The ``broad'' bandwidth of the magnetic doublon is strongly dependent on the underlying magnetic configuration. 
This is demonstrated in Fig.\ \ref{fig:W} where it is compared with the bandwidth for the case of an antiferromagnetic singlet ($S=0$). Such a ground state can easily be stabilized by adding a weak direct coupling $\dir{J} \sum_{\left<ij\right>} \v{S}_i \cdot \v{S}_j$ to the Hamiltonian ($\dir{J}=0.1$).
We find the same four types of states, but the magnetic doublon bandwidth now scales as $W \sim J^{-0.35}$. Hence, the magnetic doublon has a substantially broader dispersion if the spin system is not polarized homogeneously. 
This is intuitively clear, as it requires a bound magnon to propagate; and the antiferromagnet (but also a disordered local-moment paramagnetic state) provides a much larger phase space in terms of spin flips to assist the propagation.
We note that as compared with the conventional colossal magnetoresistance (CMR) effect, transport is alleviated by spin disorder rather than impeded. 
Hence, there is an ``inverse'' CMR effect in controlling the nonequilibrium transport properties of magnetic doublons via temperature or magnetic switching. 

\paragraph{\color{blue} Electron spectroscopy at finite filling ratio.} 
In principle, the most direct access to observing magnetic doublon excitations is given by appearance-potential spectroscopy \cite{Potthoff_1993, Fukuda_2010}, where there are two additional electrons in the final state, predominantly created at the same site. 
This is shown in the Supplemental Material \cite{suppl}. 
Here, we instead focus on $k$-resolved single-electron spectroscopy, which is a widely established experimental technique. 
For convenience we will consider electron addition, i.e. the inverse photoemission spectrum from systems with low band filling. 
This is related to standard photoemission (one-electron removal) at high band fillings via particle-hole symmetry. 

\begin{figure}
\includegraphics[width=\columnwidth]{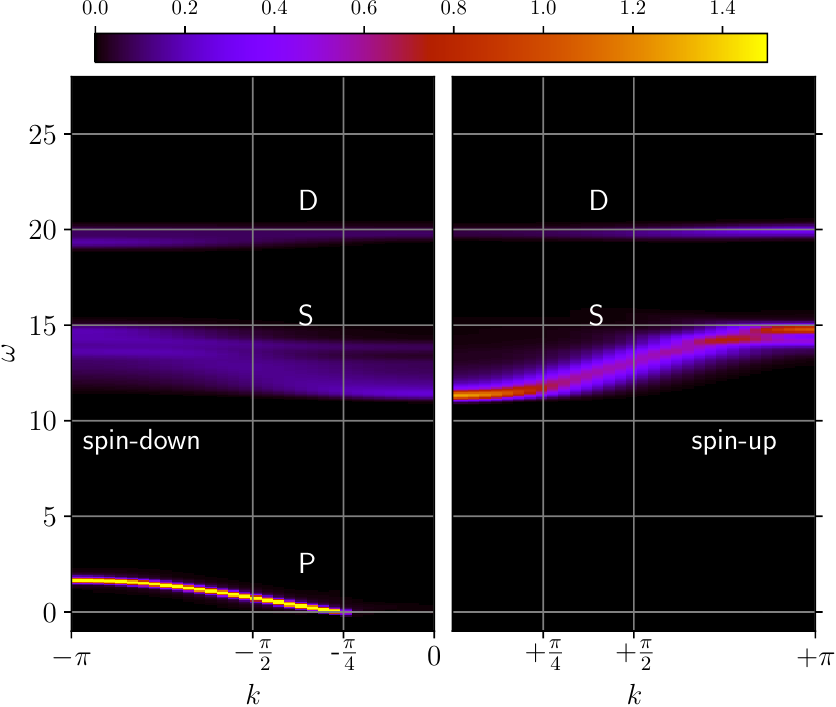}
\caption{\label{fig:IPES}
Inverse photoemission spectrum Eq.\ (\ref{eq:IPE}) as obtained by Chebyshev-expansion DMRG for $n=\nicefrac14$ ($N=16$, $L=64$), $S=(L-N)/2$ and $J=12$. 
Left: spin-down (minority). Right: spin-up (majority, shifted rigidly by $\Delta \omega \approx 11.2$ to align with the spin-down spectrum). 
Both spectra are symmetric with respect to $k=0$.
\textit{P}: polaron, \textit{S}: scattered state, \textit{D}: doublon.
}
\end{figure}

We demonstrate that the magnetic doublon can be seen as a satellite for a finite filling $n=N/L$ in the thermodynamic limit (large $N$, large $L$).
This is shown in Fig.\ \ref{fig:IPES}, which displays the spin-resolved spectrum at $n=\nicefrac14$ (one-eighth filling), given by 
\begin{equation}
A_{\sigma}\lr{k,\omega} 
= 
\langle 0 | c_{k\sigma} \delta(\omega + E_{0} - H)  c_{k\sigma}^\dagger | 0 \rangle \: , 
\label{eq:IPE}
\end{equation}
where $c_{k\sigma} = \sum_{i} e^{-ikR_{i}} c_{i\sigma} / \sqrt{L}$ and where $|0\rangle$ is the $N$-particle ground state with magnetization $M$, which is the eigenvalue of $\sum_i (S^z_i + s^{z}_i)$. 
Note that the spin-$\uparrow$ spectrum has been rigidly shifted to align with the spin-$\downarrow$ spectrum for better comparison. 
In the latter case, $\omega=0$ refers to the Fermi edge.

To calculate the spectral function, we have used a density-matrix renormalization group (DMRG) code which actually exploits the full SU(2) spin symmetry along with the U(1) charge symmetry of the model, combined with the Chebyshev polynomial technique \cite{Weisse_2006,Rausch_2016}. 
As before, we control the total spin $S$, but are able to resolve the result according to $M$ in a post-processing step from a single calculation. 

In the spin-down (minority) case, the lowest-frequency states are given by processes where the added electron forms a polaron with the magnetic substrate. 
In the strong-coupling limit, we can think of the ground state as being filled by noninteracting polarons which occupy the lowest-momenta states, so that the dispersion of the excitation (``P'') starts at $k \approx n \pi = \nicefrac\pi 4$ for $\omega=0$. 
Flipping the incident electron spin then leads to a higher scattering state (``S''), but there is also some probability to find another electron with which a doublon state can be formed (``D'').
In the spin-up (majority) case, the electron has a high probability to find empty sites (because of the low density $n$), upon which it propagates with a free dispersion of bandwidth $W=4T$ (``S''). 
However, there is again some probability to form a doublon with an electron from the ground state (``D''). 
In both cases, the energy distance of the doublon to the scattering states is about $J/2$ and to the lowest polaron states about $3J/2$, consistent with the two-electron eigenstates of Fig.\ \ref{fig:eigenstates}. 
The doublon excitation (``D'') has an extremely weak dispersion and is basically spin-independent. 
We did not attempt to resolve its internal structure, i.e. the bound magnetic doublon and the narrow continuum.

\paragraph{\color{blue} Real-time dynamics.} 
Finally, we demonstrate the stability of magnetic doublons.
To this end a local doublon-hole excitation is created suddenly at $t=0$ by applying the correlated-hopping operator $C_{i,i+1}=\sum_{\sigma}\lr{1-n_{i,-\sigma}}c_{i\sigma} c^{\dagger}_{i+1,\sigma} n_{i+1,-\sigma}$ locally to the ground state, i.e. $\ket{\Psi\lr{0}}=C_{i,i+1}\ket{0}$.
Then this state is propagated in time, $\ket{\Psi\lr{t}}=\exp\lr{-iHt}\ket{\Psi\lr{0}}$, and the total double occupancy $\avg{\tot{n}^d\lr{t}}=\sum_i\matrixel{\Psi\lr{t}}{n_i^d}{\Psi\lr{t}}$ is measured at each step.
Apart from being a general study of doublon decay in real time, this setup could also be regarded as a crude modeling of the effect of an incident photon \cite{AlHassanieh_2008}.
It is advantageous to use a correlated-hopping operator instead of just $c_{i\sigma} c^{\dagger}_{i+1,\sigma}$, as this ensures that the double occupancy will be zero on site $i$ and one on site $i+1$ independently of the initial configuration. 
We furthermore subtract the ground-state contribution of $\avg{\tot{n}^d}$ to facilitate the comparison for different values of $J$. 
Since the largest cross-section for this excitation is found at half filling, we consider the AFM singlet state ($S=0$) at $n=1$.
We first calculate the ground state of an infinite system using the variational uniform matrix-product states
(VUMPS) algorithm \cite{Zauner-Stauber_2018} with SU(2) spin and U(1) charge symmetry; then a heterogeneous section \cite{Phien_2012} is assembled where the excitation is allowed to spread \cite{Haegeman_2016}.

\begin{figure}
\includegraphics[width=\columnwidth]{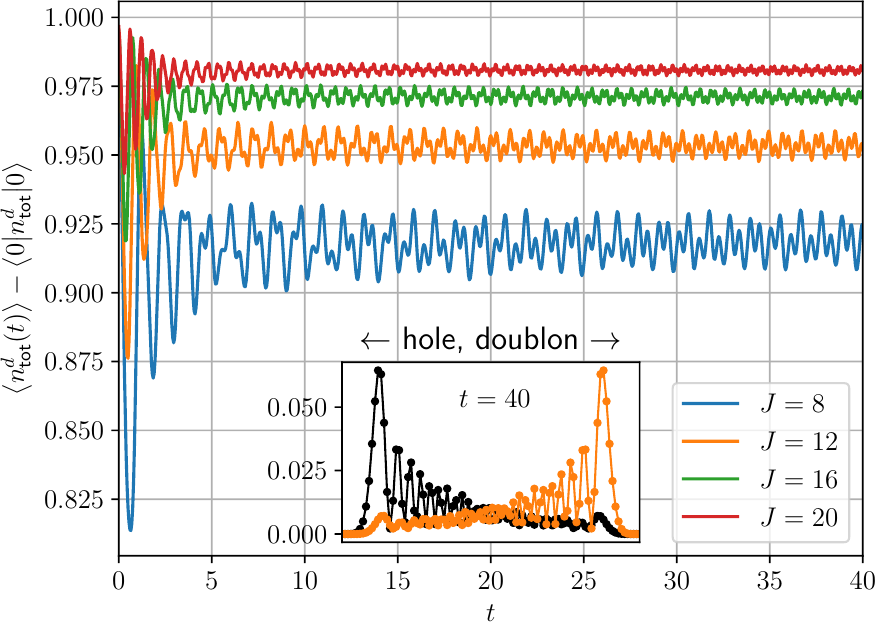}
\caption{\label{fig:dtot}
Time dependence of the total double occupancy (with ground-state value subtracted) after a doublon-hole excitation as obtained for an infinite MPS with $S=0$, $n=1$ on a heterogeneous section of $L=100$ lattice sites. Inset: real-space snapshot at time $t=40$ of the double occupancy (orange, propagating to the right) and the empty occupancy (black, propagating to the left) for $J=12$.
}
\end{figure}

We find that the magnetic doublon picture continues to hold in this regime, even as the many-body effects are strongest.
Figure \ref{fig:dtot} shows that, after a few inverse hoppings, the double occupancy settles at a constant plateau quite close to the initial value of (almost) unity without any further decay on the accessible time scale. 
The larger $J$, the closer it stays to unity.
As compared to doublon dynamics in the Hubbard model \cite{AlHassanieh_2008, Hofmann_2012, Rausch_2017}, there are important differences resulting from the different binding mechanisms: 
A Fourier analysis (not shown) indicates that there are two dominating oscillation frequencies in the Kondo case, which are roughly given by $J/2$ and $3J/2$. 
Furthermore,  in the Hubbard case the maximal velocity of the wavefront is basically given by the Fermi velocity $v_{\text{max}} \approx 2T$, while in the Kondo case we find $v_{\text{max}} \approx T$, which in fact corresponds to the polaron velocity in the strong-$J$ limit \cite{Tsunetsugu_1997}.
Thus, qualitatively, a propagating doublon excitation of an antiferromagnetic Mott insulator should be thought of as free-electron like, opposed to the polaron picture that applies to the strong-$J$ antiferromagnetic Kondo insulator. 

The inset of Fig.\ \ref{fig:dtot} shows a real-space snapshot of the double occupancy $\avg{n^d_i}$ and the density of empty sites $\avg{n^h_i}=1-\avg{n_i}+\avg{n^d_i}$ (with the ground-state value again subtracted) for $J=12$ and a late time. The two wave packets separate in opposite directions with some spread, but little charge recombination. 
This demonstrates an efficient mechanism for charge separation and thus has potential relevance for solar cell devices \cite{AlHassanieh_2008}. 

\paragraph{\color{blue} Conclusions.} 
We have introduced the notion of magnetic doublons that are formed due to magnons which bind electrons in pairs. 
Our calculations for the Kondo lattice in one dimension have not only demonstrated their existence, but also suggest a number of highly functional properties and unconventional effects to be exploited in the future. 
This includes metastable superconductivity due to Bose condensation of doublons as well as quantum distillation and efficient carrier separation.
Candidate materials could be the long-studied manganites \cite{Kubo_1972, Ishihara_1997}, rare-earth chalcogenides or Hund's metals \cite{YHK11}. 
We also note that the Kondo model in the strong-$J$ regime has been implemented on an ultracold lattice recently \cite{Riegger_2018}.
Opposed to conventional doublon excitations in the Hubbard model, magnetic doublons have a more variable dispersion which is highly sensitive to the magnetic background. 
The latter implies a colossal magnetoresistance effect, though curiously an inverse one, as the magnetic-doublon mobility decreases with increasing homogeneity of the spin system.

\begin{acknowledgments}
R.R.\ would like to thank the Japan Society for the Promotion of Science (JSPS) and the Alexander von Humboldt Foundation. 
Computations were partially performed at the Yukawa Institute for Theoretical Physics, Kyoto.
We gratefully acknowledge the support by the Deutsche Forschungsgemeinschaft within the SFB 925 (project B5), by the Cluster of Excellence ``Advanced Imaging of Matter'' EXC 2056 (project ID 390715994), as well as by JSPS KAKENHI (Grants No. JP15H05855, No. JP18H01140, No. JP18F18750, and No. JP19H01838).

\end{acknowledgments}

% \bibliography{magdoublon_suppl,litplus}


%\title{Supplemental Material for \\``Magnetic Doublon Bound States in the Kondo Lattice Model''}

%\author{Roman Rausch$^{1}$}
%\email[]{rausch.roman.72e@st.kyoto-u.ac.jp}
%\author{Michael Potthoff$^{2}$}
%\author{Norio Kawakami$^{1}$}
%
%\affiliation{$^{1}$Department of Physics, Kyoto University, Kyoto 606-8502, Japan}
%\affiliation{$^{2}$Department of Physics, University of Hamburg, Jungiusstra{\ss}e 9, D-20355 Hamburg, Germany}
%
%\date{\today}

%\maketitle

\clearpage
\section{Supplemental Material}

\subsection{Eigenstates in all sectors}

\begin{figure}[bh]
\includegraphics[width=\columnwidth]{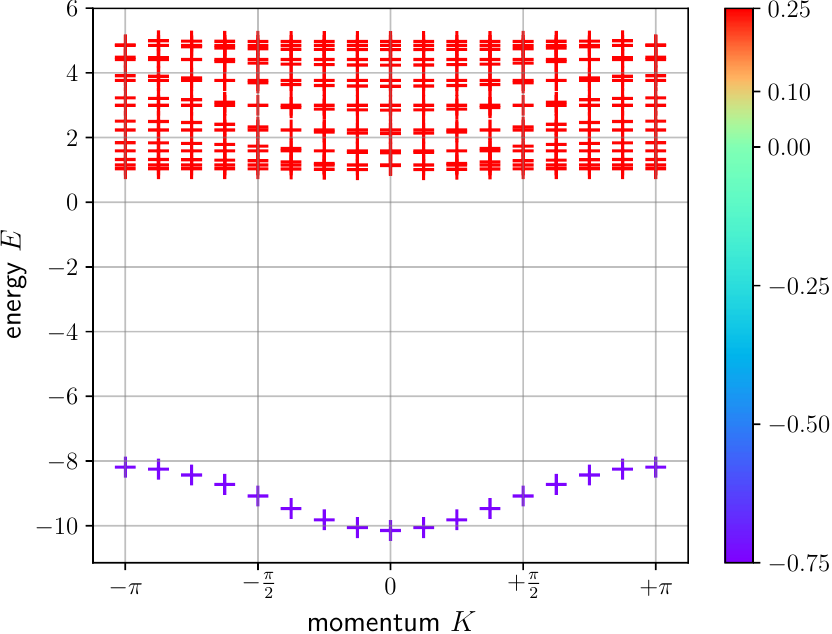}
\caption{\label{fig:eigPolaron}
Exact eigenstates for $L=16$, $N=1$, $S=\lr{L-N}/2$, $J=12$. 
The color code represents the local spin-spin correlation $\sum_i \avg{\v{S}_i\cdot\v{s}_i}$.
}
\end{figure}

For the sake of completeness, we present the eigenstates obtained by exact diagonalization in all relevant subspaces. 
Let us start with the one-electron problem. 
The ground state is ferromagnetic and lies in the sector $S=(L-1)/2$, since the kinetic energy of the electron is minimized for a fully polarized local-spin system.
Figure \ref{fig:eigPolaron} shows the numerical solution for $N=1$, $S=(L-1)/2$, which, at least to some degree, can also be computed analytically \cite{Tsunetsugu_1997, Henning_2012}. 
We find a polaron band at low energies. 
For strong $J$, the polaron basically consists of the electron forming a singlet with the localized spin on the same site. 
Delocalized polaronic eigenstates of $H$ with a dispersion of bandwidth $W\approx2T$ are obtained by a superposition of such local singlets. 
At energies of about $J$ above the polaron band, we find the scattering states, where the majority-spin electron and a magnon excitation propagate independently through the lattice. 
We conjecture that for $N>1$ but fillings $n\ll 1$, the ground state can be approximately constructed as a noninteracting sea of polarons filled up to the ``Fermi'' momenta $k=\pm n\pi$.

%%%%%%%%%%%%%%%%%%%%%%%%%%%%%%%%%%%
%%%%%%%%%%%%%%%%%%%%%%%%%%%%%%%%%%%

%%%%%%%%%%%%%%%%%%%%%%%%%%%%%%%%%%%
\begin{figure}[bh]
\includegraphics[width=\columnwidth]{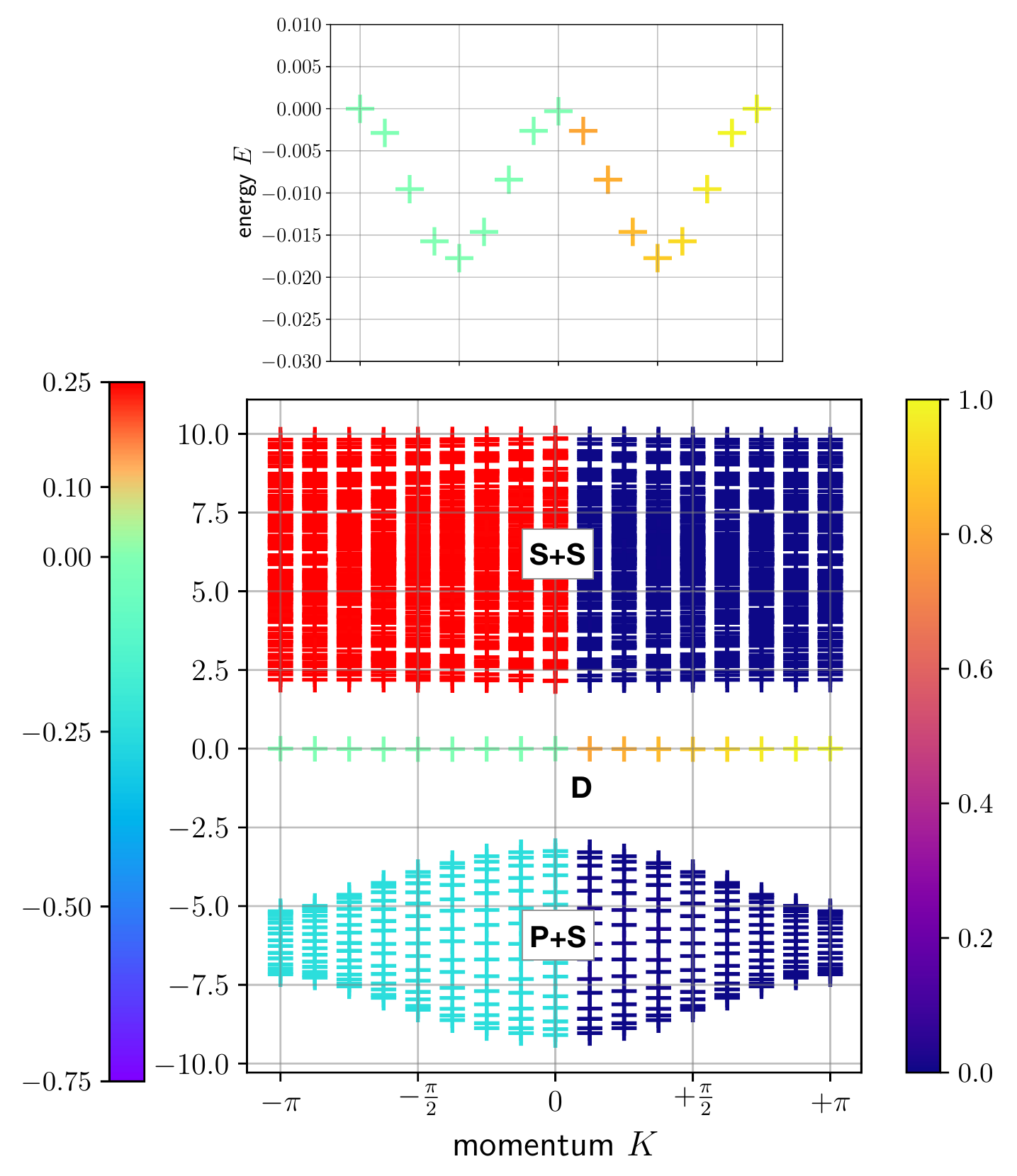}
\caption{\label{fig:eigFM}
Exact eigenstates for $L=16$, $N=2$, $S=L/2$, $J=12$. The color code represents the average local spin-spin correlation per electron $1/2 \sum_{i} \avg{\v{S}_i\cdot\v{s}_i}$ for $K\leq 0$; and the double occupancy $\sum_{i} \avg{n_{i\uparrow}n_{i\downarrow}}$ for $K>0$. 
Upper panel: close-up of the doublon band. 
See text for an explanation of the labels.}
\end{figure}
%%%%%%%%%%%%%%%%%%%%%%%%%%%%%%%%%%%

Figure \ref{fig:eigFM} presents the numerical solution of the two-electron problem in the Hilbert-space sector with $S=L/2$. 
We find the three types of eigenstates. 
\textit{P+S}: One electron forms a polaron with the localized spins, while the other one is ferromagnetically aligned to the localized spins and forms a scattering state. 
\textit{S+S}: Both electrons form scattering states.
\textit{D}: A doublon is formed by the two electrons. 
The bandwidth (see upper panel of Fig.\ \ref{fig:eigFM}) is extremely small and has minima at $k=\pm\pi/2$.

The effective model $H_{\rm eff}$ of the main paper in fact gives a vanishing band width in this sector, so that the $L$ doublon states on the $L$ sites are perfectly localized, and the doublon dispersion seen in the figure is an effect beyond the lowest-order perturbation theory.
This perfect localization is a result the restricted Hilbert space. There is only a one spin-flip deviation from the fully polarized ferromagnet which can either exist in the electronic system or be transferred to the local-spin system. In the former case, we have $\avg{\v{S}_{i}\cdot\v{S}_{j}}-\nicefrac14=0$, while in the latter case no doublon can be formed from two same-spin electrons, hence one has strictly $H_{\text{eff}} = 0$.

\begin{figure}
\includegraphics[width=\columnwidth]{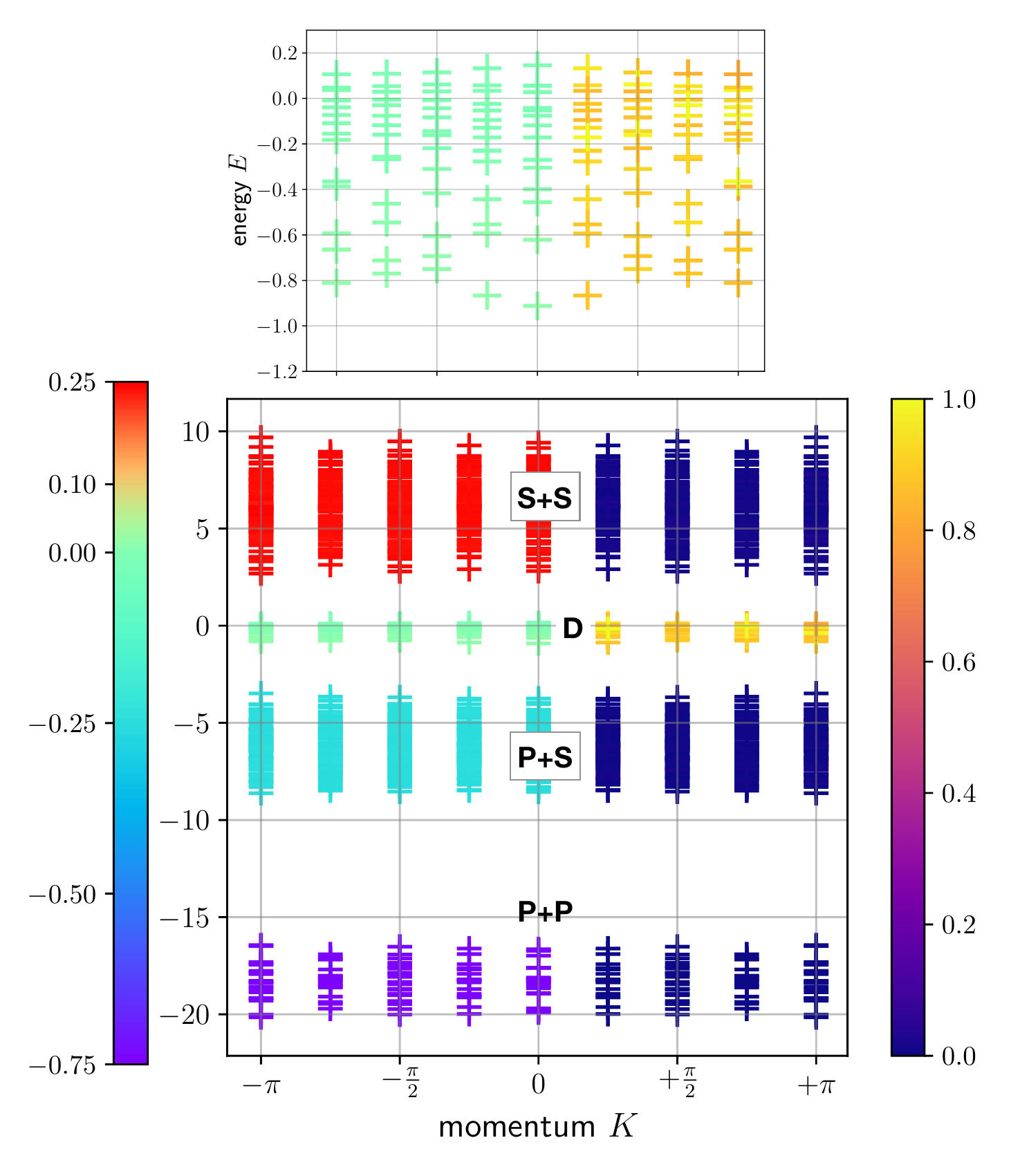}
\caption{\label{fig:eigAFM}
Exact eigenstates for $L=8$, $N=2$, $S=0$, $J=12$, $J_{\text{dir}}=0.1$. The color code represents the spin-spin correlation per electron $(1/2)\sum_{i}\avg{\v{S}_i\cdot\v{s}_i}$ for $K\leq 0$; and the double occupancy $\sum_{i} \avg{n_{i\uparrow}n_{i\downarrow}}$ for $K>0$. 
Upper panel: close-up of the doublon continuum. 
See text for an explanation of the labels.}
\end{figure}

Finally, Fig.\ \ref{fig:eigAFM} presents the solution of the two-electron problem in the largest Hilbert space sector with total spin $S=0$. 
To ensure that the ground state is antiferromagnetic and in fact lies in the $S=0$ sector, we have added a weak direct exchange term $\dir{J}\sum_{\left<ij\right>} \v{S}_i \cdot \v{S}_j$, $\dir{J}=0.1$ to the Hamiltonian. 
One observes the same types of states as in the ferromagnetic case with $S=\lr{L-2}/2$, as discussed in the main text, but the doublon bandwidth is larger by a factor of $2$-$3$. 
One now finds a doublon continuum with only a small hint of dispersion around zero momentum.

\subsection{Two-electron addition spectrum for the empty band case}

\begin{figure}
\includegraphics[width=\columnwidth]{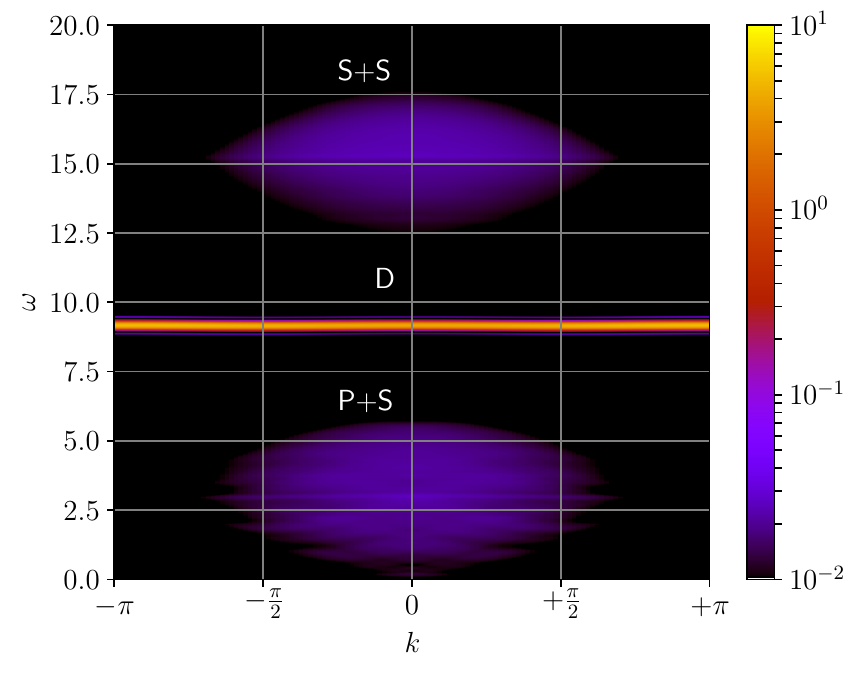}
\caption{\label{fig:APS}
Two-electron addition spectral function of the empty ferromagnetic band for $L=128$, $N=0$, $S=L/2$, $J=12$ as obtained by Chebyshev-expansion DMRG.
See Fig.\ \ref{fig:eigFM} for comparison.
}
\end{figure}

For the doublon problem in the Hubbard model, the two-electron \textit{removal} spectrum for the case of a completely filled band serves as a demonstrative and paradigmatic example \cite{Rausch_2016}.
At low fillings or for an empty band, one should rather perform a two-electron \textit{addition} spectroscopy, defined by the following spectral function:
\begin{eqnarray}
A_{\rm 2}\lr{k,\omega} 
& = &
\frac1L
\sum_n 
\norm{
\matrixel{n}{\sum_{i} e^{-ikR_{i}} c^{\dagger}_{i\downarrow}c^{\dagger}_{i\uparrow}}{0}
} \times
\nonumber \\
&\times&
\delta \lr{\omega - E_0 + E_n }
\: .
\label{eq:spectrumk}
\end{eqnarray}
$\sum_{k} A_{\rm 2}\lr{k,\omega}$ plays the role of the raw spectrum in appearance-potential spectroscopy (APS) \cite{Potthoff_1993}, a well-established technique that has gained some renewed interest recently \cite{Fukuda_2010}. 
The two-electron addition spectrum for the case of an empty band, with the localized spins aligned ferromagnetically, is displayed in Fig.\ \ref{fig:APS}. 
The subspace of the excited states $| n \rangle$ which have a non-vanishing matrix element exactly corresponds to the subspace underlying the calculations shown in Fig.\ \ref{fig:eigFM}, and hence we find the same three types of states represented. 
However, almost all of the spectral weight is concentrated on the doublon band, requiring a logarithmic scale to observe the other states at all. 
Thus, APS is in principle the optimal technique to observe the magnetic doublon. 
A major difference from the Hubbard case is the small doublon dispersion, essentially constituting a flat band whose curvature can barely be resolved.
Recall that in the Hubbard case, it scales as $U^{-1}$, while in the Kondo case, the scaling is $J^{-3}$, as discussed in the main text.
One further observes that the spectral weight of the other states disappears at the Brillouin zone edge $k\to\pm\pi$. 
This is due to the charge-SU(2) symmetry of the model, whereby an exact doublon eigenstate is created at these momenta. 
The same effect is found in the Hubbard model \cite{Rausch_2016}.

\subsection{Derivation of the effective model}

Here, we outline the derivation of the effective model for the ``broad'' dispersion of the magnetic doublon, split off from the manifold of zero-energy doublon states. 
Abbreviating $H_{\text{Kondo}} = J \sum_i \v{S}_i \cdot \v{s}_i$, we start by decomposing the lattice fermion hopping into three terms \cite{Fazekas_1999}:
\begin{equation}
H = T^0 + T^+ + T^- + H_{\text{Kondo}} \: .
\end{equation}
The first,
\begin{equation}
T^0 = -T \sum_{\left<ij\right>\sigma} \lr{ n_{i,-\sigma} c^{\dagger}_{i\sigma}c_{j\sigma} n_{j,-\sigma} + h_{i,-\sigma} c^{\dagger}_{i\sigma}c_{j\sigma} h_{j,-\sigma} + H.c.}
\end{equation}
with $h_{i\sigma} = 1-n_{i\sigma}$, is the part that conserves double occupancy, while
\begin{equation}
T^+ = -T \sum_{\left<ij\right>\sigma} \lr{ n_{i,-\sigma} c^{\dagger}_{i\sigma}c_{j\sigma} h_{j,-\sigma} + H.c.}
\end{equation}
increases it, and
\begin{equation}
T^- = -T \sum_{\left<ij\right>\sigma} \lr{ h_{i,-\sigma} c^{\dagger}_{i\sigma}c_{j\sigma} n_{j,-\sigma} + H.c.}
\end{equation}
decreases it.

We proceed with a canonical transformation, specified by the Hermitian operator $S$:
\begin{equation}
H_{\text{eff}} = e^{iS} H e^{-iS} \: .
\end{equation}
In leading order, this yields
\begin{equation}
H_{\text{eff}} = H + \commut{iS}{H} + \frac{1}{2} \commut{iS}{\commut{iS}{H}} + \ldots.
\end{equation}
The generator $S$ shall be chosen such that it cancels the terms which alter the double occupancy in first order:
\begin{equation}
\commut{iS}{H_{\text{Kondo}}} = -\lr{T^+ + T^-} \: . 
\label{eq:Scommut}
\end{equation}
This implies $S=O\lr{J^{-1}}$, and we obtain:
\begin{equation}
H_{\text{eff}} = T^0 + H_{\text{Kondo}} + \commut{iS}{T^0} + \frac{1}{2}\commut{iS}{T^+ + T^-} + O\lr{J^{-2}} \: .
\label{eq:Heff_prelim}
\end{equation}

It turns out that $\commut{iS}{T^0}=0$. Eq. (\ref{eq:Scommut}) can be solved in the eigenbasis of $H_{\text{Kondo}}$. It is a local operator for which the eigenstates $\ket{\alpha}$ can be constructed easily:
\begin{equation}
\sum_{ij} U^{\dagger}_{\alpha i} H_{\text{Kondo},ij} U_{j\beta} = D_{\alpha\beta} \delta_{\alpha\beta} \: .
\end{equation}
The generator $S$ is now explicitly given by:
\begin{equation}
S_{\alpha\beta} = i\frac{\matrixel{\alpha}{T^+ + T^-}{\beta}}{E_{\beta}-E_{\alpha}}, ~~\alpha\neq\beta \: .
\end{equation}
For $\alpha = \beta$, one simply sets $S_{\alpha\beta} \equiv 0$. 
The property $S=O\lr{J^{-1}}$ is evident from the above expression, since the eigenvalues of the Kondo term obviously scale as $E_{\alpha} \propto J$. 
Having obtained $S_{\alpha\beta}$, we have to rotate it back into the original basis:
\begin{equation}
S_{ij} = \sum_{\alpha\beta} U_{i \alpha} S_{\alpha\beta} U^{\dagger}_{\beta j} \: .
\end{equation}
We proceed by neglecting singly occupied sites, which allows us to disregard $T^0$ in Eq.\  (\ref{eq:Heff_prelim}), and arrive at the following final result:
\begin{equation}
H_{\text{eff}} = J_K' \sum_{ij}^{\rm n.n.} \lr{ d^{\dagger}_id_j + d^{\dagger}_jd_i + \lr{n^d_i-n^d_j}^2 } \cdot \lr{\v{S}_i \cdot \v{S}_j-1/4} \: .
\label{eq:Heff:suppl}
\end{equation}
Here $J_K'=8T^2/3J$. 
The doublon creation operator $d^{\dagger}_i = c^{\dagger}_{i\downarrow} c^{\dagger}_{i\uparrow}$ now becomes the creation operator of a hard-core boson, with $\commut{d_i}{d^{\dagger}_j}=\delta_{ij}$ and $\lr{d^{\dagger}_i}^2=0$; and $n^d_i=d^{\dagger}_id_i$ is the corresponding density. 
Note that in our convention every nearest-neighbor bond is counted once, so that $\sum_{ij}^{\rm n.n.} n^d_i = z/2 \sum_i n^d_i$, where $z$ is the coordination number of the lattice ($z=2$ for 1D).

We may compare it with the Hubbard result \cite{Rosch_2008}, which can be correspondingly written as:
\begin{equation}
H_{\text{eff}} = \frac{J_H'}{2} \sum_{ij}^{\rm n.n.} \lr{ d^{\dagger}_id_j + d^{\dagger}_jd_i + \lr{n^d_i-n^d_j}^2 } + U \sum_i n^d_i.
\label{eq:Heff:Hubbard}
\end{equation}

Both cases can be further simplified by mapping hard-core bosons to spins:
\begin{eqnarray}
d^{\dagger}_i &\to \lr{-1}^i T^+_i \\
d_i           &\to \lr{-1}^i T^-_i \\
n^d_i        &\to T^z_i + 1/2 \label{eq:Tz}
\end{eqnarray}
with $\commut{T_i^{z}}{T_j^{\pm}}=\pm T_i^{\pm} \delta_{ij}$ and $\commut{T^+_i}{T^-_j}=2T^z_i\delta_{ij}$. 
This leads to a bi-quadratic ferromagnetic spin model:
\begin{equation}
H_{\text{eff}} = -2J_K' \sum_{ij}^{\rm n.n.} \lr{\v{T}_i \cdot \v{T}_j-1/4} \cdot \lr{\v{S}_i \cdot \v{S}_j-1/4}.
\end{equation}

\begin{figure}
\includegraphics[width=\columnwidth]{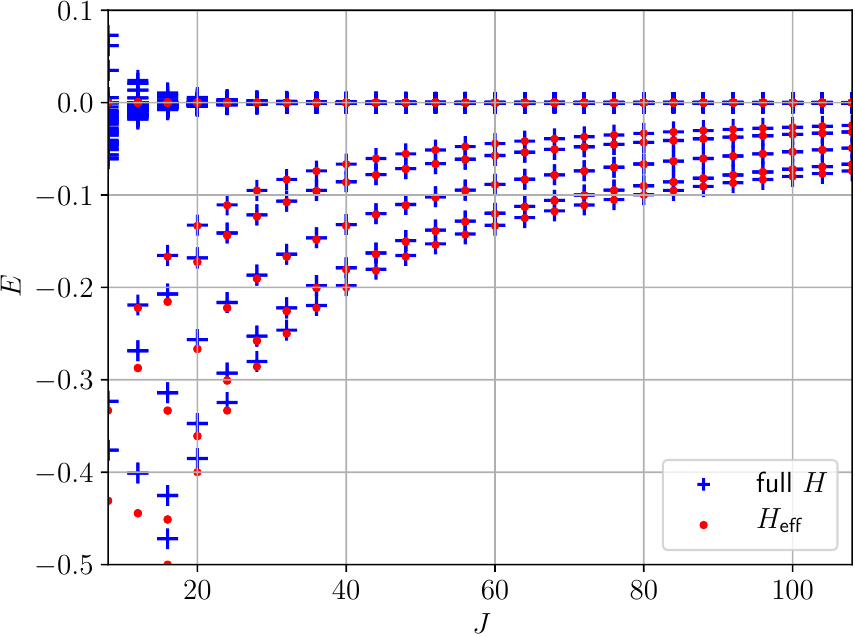}
\caption{\label{fig:Heff}
Eigenstates for $L=8$, $N=2$, $S=(L-N)/2$ as a function of $J$ within the doublon band (blue crosses) compared with the effective model Eq.\ (\ref{eq:Heff:suppl}) (red dots).
}
\end{figure}

The validity of the effective model is demonstrated in Fig.\ \ref{fig:Heff} in the subspace of a single doublon and a single magnon. 
There are $L/2+1$ distinct $k$-values in the broad magnetic doublon band, which agree with the effective model for strong $J$. 
States where the doublon and the magnon are not nearest neighbors give exactly $E=0$ in the effective model. 
In the full model, this degeneracy seems to be lifted in third order, resulting in the narrow doublon continuum with a bandwidth $W \sim J^{-3}$.

\begin{figure}
\includegraphics[width=\columnwidth]{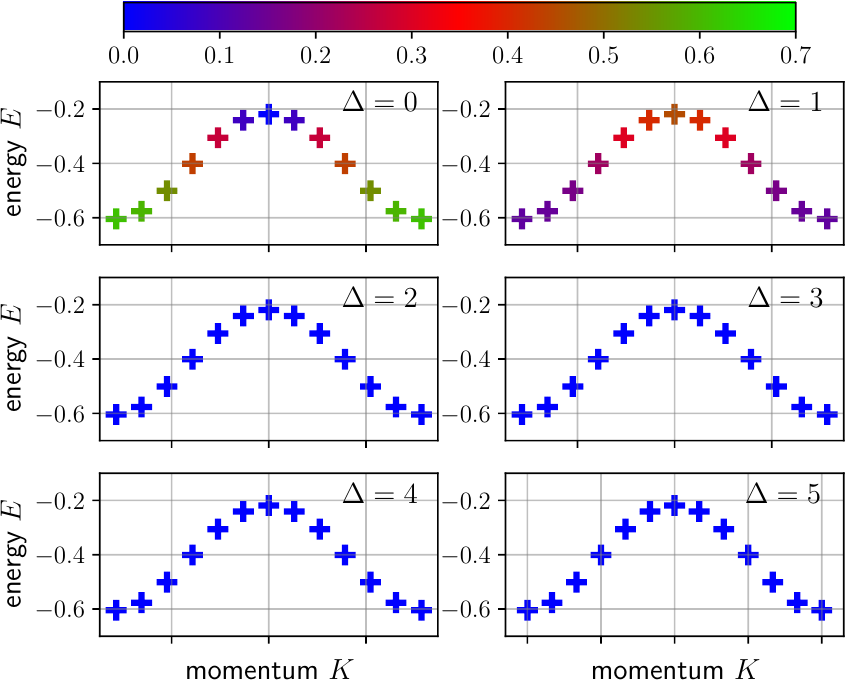}
\caption{\label{fig:nmnd}
Magnon-doublon correlation function Eq. (\ref{eq:nmnd}) calculated for $L=12$, $N=2$, $S=(L-N)/2$ and different magnon-doublon separations within the doublon band.
}
\end{figure}

Another test one can do is to compare whether the effect of only nearest-neighbor correlation holds up in the full model. For that, one can look at the following correlation function:
\begin{equation}
C_{md}\lr{\Delta} = \avg{n^m_in^d_{i+\Delta}} = \avg{\lr{1/2-S^z_i}n_{i+\Delta,\uparrow}n_{i+\Delta,\downarrow}},
\label{eq:nmnd}
\end{equation}
where $n^m_i:=1/2-S^z_i$ is the density operator of the down-spins reinterpreted as fermions via the Jordan-Wigner transformation. This correlation function is thus the density-density correlation between magnons and doublons, with $\Delta$ being the separation of the two. The result is displayed in Fig. \ref{fig:nmnd}.
And indeed, one observes a sudden drop to zero for a separation of $\Delta \geq 2$, i.e. beyond nearest neighbors of the chain. For the total momentum $K\to\pm\pi$, the correlation is highest at $\Delta=0$, so that the binding actually seems to be favored at high kinetic energies. Note that for $K\to0$ the narrow doublon continuum and the broad doublon band are closest to each other (see main text). Therefore, we may expect that hybridization effects between the two appear which go beyond the effective model.
 
\subsection{Technical information on the DMRG calculation}

For the dynamics of the Kondo insulator at half filling we have to start from the ground state, which is rather simple: For $J\to\infty$ it is a product state with zero entanglement entropy which consists out of local singlets. For large but finite $J$, the entanglement is still very small, so that the ground state converges easily. After the local excitation we follow the excited wavepacket adaptively through the time evolution using an algorithm based on the time-dependent variational principle (TDVP) \cite{Haegeman_2016} with a time step of $\delta t=0.05$. The maximal SU(2)-invariant bond dimension $\chi_{\text{SU(2)}}$ reaches about $70-80$ (corresponding to $\chi \approx 160-190$ without exploiting the SU(2) symmetry) and the truncation error is around $10^{-7}$ to $10^{-8}$.

The inverse photoemission at $\nicefrac18$ filling is a considerably harder problem. We calculate about $1300$ Chebyshev moments, which corresponds to an energy resolution of $\delta E/T\approx0.2$. The distance tolerance between the Chebyshev recursion vectors is set to $10^{-4}$. To achieve it, $\chi_{\text{SU(2)}}$ reaches $435$ during the calculation. 

\end{document}